\documentclass{ws-procs975x65}

\begin{document}



\title{ SECONDARY PERTURBATION EFFECTS IN KEPLERIAN ACCRETION DISKS: ELLIPTICAL INSTABILITY
}

\author{BANIBRATA MUKHOPADHYAY
}

\address{Department of Physics,
Indian Institute of Science, 
Bangalore-560012, India 
\email{bm@physics.iisc.ernet.in}}

%


\def\lsim{\lower.5ex\hbox{$\; \buildrel < \over \sim \;$}}
\def\gsim{\lower.5ex\hbox{$\; \buildrel > \over \sim \;$}}

\def\ch{\lower-0.55ex\hbox{--}\kern-0.55em{\lower0.15ex\hbox{$h$}}}
\def\lh{\lower-0.55ex\hbox{--}\kern-0.55em{\lower0.15ex\hbox{$\lambda$}}}

\begin{abstract}
Origin of turbulence in cold accretion disks, particularly in 3D, which 
is expected to be hydrodynamic but not magnetohydrodynamic, is a big puzzle. 
While the flow must exhibit some turbulence in support of the transfer of mass 
inward and angular momentum outward, according to the linear perturbation 
theory it should always be stable. We demonstrate that the 3D secondary 
disturbance to the primarily perturbed disk which exhibits elliptical vortices 
into the system solves the problem. This result is essentially
applicable to the outer region of accretion disks in active galactic nuclei 
where the gas is significantly cold and neutral in charge and the magnetic 
Reynolds number is smaller than $10^4$.
\end{abstract}

\bodymatter

\section{Introduction}\label{intro}

\enlargethispage*{6pt}

Despite much effort devoted, the origin of hydrodynamic
turbulence in Keplerian accretion disks is still poorly understood. 
This is essentially important for accretion disks
around quiescent cataclysmic variables,
proto-planetary and star-forming disks, and
the outer regions of disks in active galactic nuclei \cite{gm}.

A Keplerian accretion disk flow having a very low
molecular viscosity must generate turbulence and successively
diffusive viscosity to support transfer of mass inward.
However, theoretically this flow never exhibits any unstable mode.
On the other hand, laboratory
experiments of Taylor-Couette systems, which are similar to
Keplerian disks, seem to indicate that although the Coriolis force
delays the onset of turbulence, the flow is ultimately unstable
to turbulence \cite{richard2001}. However, some other experiments
say against it \cite{ji}. We believe that not finding hydrodynamic turbulence
is due to their choice of large aspect ratio and small Reynolds number.

Various kinds of secondary instability, such as the elliptical
instability, are widely discussed as a possible route to
self-sustained turbulence in linearly perturbed shear flows (see,
e.g. \cite{pier86}). These effects
have been proposed as a generic mechanism
for the breakdown of many two-dimensional high Reynolds number
flows. However, such effects have not been discussed properly
in literatures for rotating Keplerian flows.

Therefore, we plan to show that the three-dimensional secondary
perturbation can generate large growth in the flow time scale and presumably trigger turbulence
in the Keplerian disk. Unlike the two-dimensional transient growth studied earlier
\cite{man} which were shown to be killed immediately in presence of vertical structure,
in the present case we demonstrate essentially three-dimensional
growth. Possibility of large growth in shear flows
{\it with rotation} by a three-dimensional perturbation opens a new window to explain
hydrodynamic turbulence which has an important implication not only in astrophysics
but also in general physics and fluid dynamics.

\section{Perturbation equations }\label{prob}

We consider a small section of the Keplerian disk in the shearing box approximation \cite{man}
with the unperturbed velocity $\vec{U}_p=(0,-x,0)$.
We also assume that the incompressible flow extends from $x=-1$
to $+1$ with no-slip
boundary conditions and is unbounded with periodic boundary condition
along $y$ and $z$. Therefore the linearized Navier-Stokes equations
for a 2D {\it primary perturbation} such that $U^p_x\rightarrow w_x(x,y,z,t)$, 
$U^p_y\rightarrow U^p(x)+w_y(x,y,z,t)$, and pressure 
$\bar{p}\rightarrow \bar{p}+p(x,y,z,t)$ are given by
\begin{eqnarray}
{dw_x\over dt} = 2\Omega w_y - {\partial {p}\over \partial x}
+\frac{1}{R_e} {\nabla}^2 w_x,\,\,\,
{dw_y\over dt} = \Omega (q-2) w_x - {\partial {p}\over \partial y}
+\frac{1}{R_e}{\nabla}^2 w_y,
\label{ymmtm}
\end{eqnarray}
with the equation of continuity
${\partial w_x/\partial x} + {\partial w_y/\partial y}
= 0$,
where the angular frequency $\Omega=1/q$ and $q=3/2$ for a Keplerian disk.
When the Reynolds number $R_e$ is very large, the solution of above equations
is
\begin{equation}
w_x=\zeta\frac{k_y}{l^2}\sin(k_xx+k_yy),\,\,w_y=-
\zeta\frac{k_x}{l^2}\sin(k_xx+k_yy),\,\,l^2=k_x^2+k_y^2,
\label{solpr}
\end{equation}
where $\zeta$ is the amplitude of vorticity perturbation.
Under this {\it primary perturbation}, the flow velocity and pressure
modify to
\begin{eqnarray}
\vec{U}&=&\vec{U}^p+\vec{w}=(w_x,-x+w_y,0)={\bf A}.\vec{d},\,\,\, p^\prime=\bar{p}+p(x,y,z,t),
\label{primper}
\end{eqnarray}
where $\vec{d}$ is the position vector and $\bf A$ is a tensor of rank $2$.

Now we concentrate on a further small patch of the primarily perturbed flow such that
$x<<1/k_x$, $y<<1/k_y$.
Then we consider a {\it secondary perturbation} to this flow such that
$\vec{U}\rightarrow \vec{U}+\vec{u}$ and $p^\prime\rightarrow p^\prime +\tilde{p}$
with
\begin{equation}
(u_{i},\tilde{p})=(v_{i}(t),p(t)) \exp(ik_m(t) x^m),\,\,\,i,m=1,2,3.
\label{pet}
\end{equation}
Therefore, we obtain the evolution of a linearized {\it secondary perturbation}
\begin{equation}
\dot{v}_j+A_j^k\,v_k+2\,\epsilon_{mkj}\Omega^m v^k=-ip\,k_j
-\frac{v_j}{R_e}\,k^2,
\label{perteq}
\end{equation}
\begin{equation}
\dot{k}_j=-(A^m_j)^T\,k_m,\,\,\,\,\,
k^n\dot{v}_n=k^m\,A^n_m\,v_n,\,\,\,\,k^2=k_mk^m.
\label{keq}
\end{equation}

\section{Solution}\label{sol}

We specifically concentrate on the flow having low viscosity. Therefore, 
the general solution \cite{m06} of eqn. (\ref{perteq})
can be written as a linear superposition of the Floquet modes
\begin{equation}
v_i(t)=\exp(\sigma\,t)\,f_i(\phi),
\label{flo}
\end{equation}
where $\phi=\varpi\,t$, $f_i(\phi)$ is a periodic function having
time-period $T=2\pi/\varpi$, and $\sigma$ is the Floquet exponent
which is different at different $\epsilon=(k_x/l)^2$.
Clearly, if $\sigma$ is positive then
the system is unstable and plausibly turbulent.

When the secondary perturbation evolves much rapidly than the primary one
and $k_i(0)=(0,0,1)$, $\sigma$ has an analytic solution given by
\begin{equation}
\sigma=\sqrt{\zeta\epsilon-(2\Omega_3-1)(2\Omega_3-\zeta)}.
\label{sigcon}
\end{equation}
Clearly, for a Keplerian disk 
$\sigma=\sqrt{\zeta\epsilon-(4-3\zeta)/9}$.
Therefore, a Keplerian flow 
is hydrodynamically unstable under a vertical secondary perturbation if $\zeta>1/3$.
For other perturbations $\sigma$ can be computed numerically described in detail
elsewhere \cite{m06}.

We now plan to
quantify this by computing the corresponding turbulent viscosity.
For the isotropic disk fluid, the turbulent viscosity
$\nu_t=\alpha c_s h$ \cite{ss73}, where $c_s$ and $h$ are local sound speed and disk thickness respectively. 
On the other hand, shearing stress $T_{xy}=<u_1 u_2>=-\nu_t q\Omega$. Therefore, 
we obtain the {\it Shakura-Sunyaev viscosity} parameter at a disk radius $r$
\begin{eqnarray}
\alpha=-\frac{T_{xy}}{q\Omega^2\left(\frac{h}{r}\right)^3 Mr^2},\,\,\,{\rm where}\,\,\,
M=\frac{\Omega x}{c_s}.
\label{alf}
\end{eqnarray} 
For a disk with $\zeta=0.35$, $k_x=3$, $k_y=0.7$, a vertical secondary perturbation evolving
for time $t_m=10$ at a disk radius $r=15$ and thickness $h(r)/r=0.05$, 
$\alpha\sim 0.02$ (the detailed discussion is reported elsewhere \cite{sm}). 
This is interesting as $\alpha$ due to MRI computed by
previous authors \cite{mri} is similar order of magnitude. 

\section{Discussion}\label{dis}

Above results verify that 
the three-dimensional growth rate due to
the secondary perturbation in a Keplerian disk
could be real and positive and corresponding growth may be
exponential and significant enough to trigger elliptical instability. 
This eventually may trigger non-linearity and then plausible turbulence in the flow time scale.
As this growth is the result of a three-dimensional perturbation,
underlying perturbation effect should survive even in presence
of viscosity. We also see that the corresponding viscosity $\alpha$ to transport
matter inward and angular momentum outward is significant and comparable to that
due to MRI.






\end{document}